\documentclass[12pt]{article}

\usepackage{a4wide}
\usepackage{amsmath}
\usepackage{amsfonts}
\usepackage{amssymb}
\usepackage{cite}

\makeatletter

\@addtoreset{equation}{section}
\makeatother

\def\Ric{\mathrm{Ric}}
\newcommand{\cd}{\!\cdot\!}
\def\half{\frac12}

\def \be { \begin{equation} }
\def \ee { \end{equation} }
\def \beal#1 { \begin{align} #1 \end{align} }
\def \nn {\notag\\}

\begin{document}

\begin{titlepage}

\title{
\vspace{-2cm}
       \vspace{1.5cm}
       Gravitational Equilibrium with Steady Flow \\ 
       and Relativistic Local Thermodynamics
       \vspace{1.cm}
}
\author{
Shuichi Yokoyama\thanks{syr18046\textcircled{a}fc.ritsumei.ac.jp},\; 
\\[25pt] 
{\normalsize\it Department of Physical Sciences, College of Science and Engineering,} \\
{\normalsize\it Ritsumeikan University, Shiga 525-0058, Japan}
}

\date{}

\maketitle

\thispagestyle{empty}


\begin{abstract}
\vspace{0.3cm}
\normalsize

A relativistic self-gravitating equilibrium system with steady flow as well as spherical symmetry is discovered. The energy-momentum tensor contains the contribution of a current related to the flow and the metric tensor does an off-diagonal component to balance with the flow momentum. 
The presence of the off-diagonal component of the metric implies the radial motion of the reference frame, which gives rise to a problem how the relativistic effect is included in thermodynamic observables for such a general relativistic system.
This problem is solved by taking an instantaneously rest frame in which geometric thermodynamic observables read as previously and giving them the special relativistic effect emerged from the inverse transformation to the original frame pointwise. 
The solution of the thermodynamic observables in accord with the laws of  thermodynamics and the theory of relativity is presented. 
Finally the relativistic structure equations for the equilibrium are derived, from which the general relativistic Poisson equation as well as the heat conduction one are developed exactly. 

\end{abstract}
\end{titlepage}

\section{Introduction} 
\label{Introduction} 

The activity of the Sun is indispensable for all lives on the Earth and has attracted a great deal of attention from common people and astrophysicists. 
In the past, only superficial information of the solar activity could be observed, but nowadays even its inside information has been gradually uncovered by highly developed observational instruments and satellites. (See \cite{cooper2013our} for instance.) 
According to observations accumulated so far, the interior of the Sun is approximately separated into several parts by its physical property such that 
the convection zone starts from the surface to the tachocline, across which the radiation zone starts and extends to the core in the deepest region. Energy is produced by nuclear reaction in the core and transported steadily through the radiation zone to the surface as the sunshine.  

A motivation of the study in this paper is to enhance the theoretical understanding of such energy transport phenomena inside the Sun. Since the energy transport occurs approximately in a steady manner as observed every day, there must exist a gravitational solution to describe such equilibrium of fluid with steady radial energy flow.  
A goal of this study is to find structure equations for such gravitational equilibrium in the framework of general relativity. One of them is the equation of balance between attractive force of gravity and repulsive force of matter, which can be viewed as an extension of the so-called Tolman-Oppenheimer-Volkov (TOV) equation given in \cite{PhysRev.55.374} including the contribution of radial steady flow. 

The construction of such a gravitational solution physically motivated by the solar internal dynamics is also theoretically important to understand how local thermodynamics and the theory of relativity interplay with each other in gravitational equilibrium. 
The study of relativistic thermodynamics was initiated soon after the theory of special relativity was made public \cite{planckRT,Hasenohrl1907,einstein1907uber}, in which the transformation rule of thermodynamic observables under Lorentz transformation was investigated and it was argued that the entropy of the system be invariant thereunder while the temperature and the heat quantity be transformed similarly to that of length inducing the so-called Fitzgerald-Lorentz contraction. 
Later it was claimed that the correct transformation rule of the temperature and the heat quantity is the same as that of time leading to its delay  \cite{Ott1963LorentzTransformationDW,Arzelis1965TransformationRD,møller1967relativistic}.  
On the other hand, thermodynamics with general relativistic effect taken into account was investigated in hydrostatic equilibrium with spherical symmetry \cite{PhysRev.35.904,PhysRev.36.1791}, in which the local temperature was discovered to be equivalent to the gravitational potential by the so-called Tolman relation. Subsequently the study of general relativistic local thermodynamics was developed in \cite{PhysRev.58.919,Taub:1948zz,RevModPhys.21.531,landau} and further by taking into account angular momentum of a rotating perfect fluid \cite{1967ApJ...150.1005H,1970ApJ...161..111H,1970ApJ...162...71B,HORWITZ1973301}.

Although these researches have enhanced the understanding of relativistic thermodynamics, there still remains an unsolved issue how macroscopic quantities are defined in accord with the laws of thermodynamics as well as the theory of relativity.
This issue was addressed by recent research of the author \cite{Yokoyama:2023nld}, in which the solution was given in the classic hydrostatic equilibrium system with spherical symmetry by refining a method proposed in \cite{Aoki:2020nzm} and constructing the entropy current as a conserved non-Noether current. As evidence it was proved that the constructed entropy density satisfies the local thermodynamic relations rigorously.%
\footnote{ 
What is refined in \cite{Yokoyama:2023nld} is how to find a non-trivial solution of a differential equation to construct a non-Noether conserved current to describe the entropy of the system. To achieve this, it is required to assume the existence of a local solution with nonzero radial flow velocity. This assumption is reasonable, matching physical intuition as discussed above, and indeed holds as shown below. 
} 
Another motivation of the study is to make progress in this direction of research and to uncover any deep connection between relativistic self-gravitating equilibrium and local thermodynamics. 
In this context, turning on steady flow enriches the interplay between them. For example, in the geometric side, a natural reference frame of coordinates is a radially moving one to be in balance with the flow momentum, which gives rise to a problem how the relativistic effect is taken into account in thermodynamic observables. In the matter side, the flow describes the particle motion, from which a problem arises how the chemical potential for the particle movement is evaluated. Another goal of this study is to determine the thermodynamic observables consistent with the theory of relativity and the laws of thermodynamics. 

A new method to reach the goals is invented and the solution of the problems raised above is presented below. 

\section{Self-gravitating equilibrium with steady flow} 
\label{RHE}

To the end of the goals, consider a general spherically symmetric system of fluid in self-gravitating equilibrium with the space-time dimension $d$ more than two.
If such a system of fluid is approximately static, then it is well described by a perfect fluid. 
However, if the fluid also flows in a steady manner, then it is a priori unknown whether a perfect fluid suits to describe such a situation. What are suitable forms of the energy-stress tensor and the line element to describe this system? 

To answer this question, suppose a situation where fluid is constituted by particles. 
Then the nonzero flow velocity means the existence of particle motion inducing the nonzero particle number current. 
This contributes to the energy-stress tensor and it must be included in accord with general covariance.
Its minimal form is  
\be 
T^{\lambda\sigma} = (\mathring \rho + \mathring p) u^\lambda u^\sigma + \mathring p g^{\lambda\sigma} + j^\lambda u^\sigma+ j^\sigma u^\lambda, 
\label{EST}
\ee
where $u^\lambda, j^\lambda$ are the fluid velocity, the particle number current, respectively, and  $\mathring \rho, \mathring p$ are energy density and pressure of the fluid in the case without particle motion. 
Although it is not necessary to attribute the current $j^\lambda$ to the particle motion, the physical interpretation is helpful for understanding and assumed in what follows. 
Accordingly the metric tensor to describe gravitational force in balance with the energy-momentum tensor must be modified. 
If the system does not break spherical symmetry by the current flowing in the radial direction, then the metric ansatz suitable to describe such a situation will be an extension of that for the perfect fluid compatible with the Gullstrand-Painlev\'e metric, which is also used to describe classic black holes to view the geometry as certain (float, rain or river) flow \cite{Gautreau:1978zz,kraus1994simplestationarylineelement,Visser_1998,taylor2000exploring,Martel_2001
}, (see also \cite{Yokoyama:2021yzf}). Such a line element is given by
\beal{
g_{\lambda\sigma}dx^\lambda dx^\sigma
=& - e^A (dt)^2 + e^B (dr +C dt)^2 +r^2\tilde g_{ij} dx^idx^j, 
\label{LineElement}
}
where $\tilde g_{ij}$ is the metric for a $(d-2)$-dimensional internal (Einstein) manifold whose Ricci tensor is given by $(d-3)\tilde g_{ij}$ and $A, B, C$ are functions only of $r$.
One of the newly introduced variable $C$ in an off-diagonal component of the metric tensor is proportional to the radial component of the velocity of the radially moving frame. 
To see it, consider a test particle with unit mass which can move only in the radial direction. The radial velocity of this particle observed in the frame is given by $e^{\frac{B-A}2}(\frac{d r}{d t}+C)$, in which the first term is the velocity in reference to the rest frame and the second one describes the relative velocity to the current frame. Therefore the radial velocity of the current moving frame is given by $\beta  = -e^{\frac{B-A}2}C$. 
Indeed, consider a new coordinate system $(\check t, \check r, \check x^i)$ defined by $\check t=t, \check r=r+ C t, \check x^i=x^i$. In this frame the radial velocity of the test particle is observed as $e^{\frac{B-A}2}\frac{d\check r}{d\check t}$ at $t=0$, when the line element becomes $ - e^A (d\check t)^2 + e^B d\check r^2 +\check r^2\tilde g_{ij} d\check x^id\check x^j$ formally identical to the previous one of the hydrostatic equilibrium system. This happens at and only at $t=0$. Such an instantaneously rest frame becomes important to evaluate a geometric class of thermodynamic observables as seen below. 
 
The variables newly introduced are not independent but related to each other by equations of equilibrium state. 
Relativistic equations of motion for fluid are obtained from the covariant conservation or continuity equation for the energy-stress tensor, $\nabla_\lambda T^{\lambda}\!_\sigma =0$, which can be rewritten as $ \frac1{\sqrt{-g}}\partial_r (\sqrt{-g} T^r\!_\sigma) = \Gamma_{\lambda\sigma}^\gamma T^\lambda\!_\gamma$. 
For $\sigma=i$, it is trivially satisfied, since $T^a\!_i$ and $\Gamma^a_{bi}$ both vanish with $a=0,r$. 
For $\sigma=0$, the right hand side vanishes, so it can be solved as $T^r\!_0 = c_\circ/\sqrt{-g}$, where $c_\circ$ is an integration constant. The integration constant can be determined by the Einstein equation, which imposes $c_\circ$ to vanish since the corresponding component of the Einstein tensor does. (See appendix \ref{Ricci}.) This leads to a constraint for the flow variables such that 
$j^r =- (\mathring p + \mathring \rho +\frac{j_0}{u_0})u^r. $
This constraint equation enables one to rewrite the other components of the energy-stress tensor as
\beal{
T^0\!_0 =& -\mathring\rho + 2 u\cd j - \varpi (u^r)^2, \quad 
T^0\!_r = - e^{-A}C\varpi, \quad 
T^r\!_r =\mathring p+ \varpi (u^r)^2, \quad 
T^i\!_j =\mathring p\delta^i_j, 
\label{EST}
}
where $u\cd j:=u^\lambda j_\lambda$ and $\varpi :=  (2 u\cdot j  -\mathring p -\mathring\rho ) /(e^{-B} -e ^{-A}C^2+2 (u^r)^2)$.
A more non-trivial fluid equation is obtained from the continuity equation with $\sigma=r$ as
\beal{
 (\mathring p+ \varpi (u^r)^2)' + \frac{ (d-2) }{r} \varpi (u^r)^2 =&-\half  ( \mathring\rho + \mathring p - 2  j\cd u  + 2 \varpi (u^r)^2 )  (\log(e^{A} - e^{B}C^2) )', 
\label{ii}
}
where $\frac1{\sqrt{-g}}\partial_r (\sqrt{-g} T^r\!_r) =(T^r\!_r)'+ ( \frac{d-2}r  + \frac{(A+B)'}2 ) T^r\!_r,  \Gamma_{\lambda r}^\gamma T^\lambda\!_\gamma =\frac{1}{2}\{ \left(e^{-A } g_{00}'+B '+A '\right)T^r\!_r- e^{-A } g_{00}' T^0\!_0 \}+ (d-2)\mathring p/r $, and \eqref{EST} were used. 

Other equations of equilibrium state are obtained from the Einstein equation, $\Ric^\lambda\!_\sigma -\frac12 \delta^\lambda_\sigma R = \kappa^2 T^\lambda\!_\sigma$, where $\kappa^2$ is the gravity coupling constant given by $8\pi G$ in $d=4$ with $G$ the Newton constant. 
The component with $(\lambda,\sigma)=(0,r)$ gives $(d-2)e^{-A}C(B+A)'/(2r) = \kappa^2(-e^{-A}C \varpi)$, that is 
\be 
\frac{d-2}{2r}(B+A)' = -\kappa^2 \varpi,  
\label{0r}
\ee
because $C\not=0$.  
Under the equation \eqref{0r}, the $(\lambda,\sigma)=(r,r)$ and $(\lambda,\sigma)=(0,0)$ components of the Einstein equation are equivalent, which reduces to  
\beal{
(d-2) \{(d -3) \left(e^{-B} - e^{-A}C^2 -1\right)+r e^{-B-A} (e^{A} - e^{B}C^2)'\}=2 r^2\kappa^2(\mathring p+ \varpi (u^r)^2). 
\label{rr} 
}

\eqref{ii}, \eqref{0r}, \eqref{rr} are the relativistic fluid equations of motion in self-gravitating equilibrium with steady flow and spherical symmetry.
Note that turning off the flow variables reduces to the previous result of hydrostatic equilibrium of a perfect fluid. (See \cite{Aoki:2020prb} for the result with general $d$.)
These must encode information of equilibrium such as local thermodynamic relations and structure equations of self-gravitating equilibrium, which is to be explored in the next section.   

\section{General relativistic structure equations and local thermodynamics} 
\label{GRLTE}

Now explore local thermodynamics and structure equations in the self-gravitating equilibrium fluid system with steady flow and spherical symmetry.   
To the end, it is required to specify the correct forms of thermodynamic observables. What are they?

To answer this question, it is helpful to first recall the previous successful case without energy flow investigated in \cite{Yokoyama:2023nld}, in which the entropy density in the rest frame denoted by $\mathring s$ was constructed as the charge density of a non-Noether conserved current (not as a scalar) by the method proposed in \cite{Aoki:2020nzm}. Then the entropy of the system is obtained by the volume integration of the entropy density, and thus the entropy is invariant under general coordinate transformation as argued in \cite{Aoki:2020nzm}. This transformation rule of entropy is consistent with the traditional one by Planck \cite{planckRT}. 
The form of the thermodynamic local infinitesimal volume was unconventionally specified by the measure of the volume element at a constant time slice, $\mathring v=e^{\mathring B/2} r^{d-2}\sqrt{\tilde g}$, where the line element was given by $ - e^{\mathring A} (dt)^2 + e^{\mathring B} dr^2 +r^2\tilde g_{ij} dx^idx^j$. This and the local Euler's relation impose the local temperature to be the so-called Tolman temperature \cite{PhysRev.35.904}, $\mathring T= T_\circ e^{-\mathring A/2}$, where $T_\circ$ is an integration constant.%
\footnote{This integration constant describes the arbitrariness of the reference point of the gravitational potential and that of the absolute temperature.} 
This is consistent with the transformation rule expected from the local Euler's relation, $\mathring T\mathring s =\mathring u +\mathring p\mathring v$, where $\mathring u=\mathring \rho\mathring v$ is the internal energy density, since the energy density $\mathring \rho$ and the pressure $\mathring p$ are given as scalars in the hydrostatic equilibrium system. In this formulation the first law of (local) thermodynamics was proven as $\mathring Td\mathring s =d\mathring u +\mathring pd\mathring v$. Note that the flow variables are all turned off in this argument. 

The above result in the spherically symmetric hydrostatic equilibrium holds for any point of spacetime. Therefore the argument to identify the local temperature and the thermodynamic infinitesimal volume is expected to hold for a gravitational equilibrium system observed in any rest frame of coordinates since they are defined only by employing geometric quantities.
Such thermodynamic observable quantities are referred to as geometric thermodynamic observables. The local temperature $\check T$ and the thermodynamic infinitesimal volume $\check v$ observed in the instantaneously rest frame with the line element $ - e^A (d\check t)^2 + e^B d\check r^2 +\check r^2\tilde g_{ij} d\check x^id\check x^j$ are determined as $\check T= T_\circ e^{-A/2}$, $\check v=e^{B/2} \check r^{d-2}\sqrt{\tilde g}$ as in the hydrostatic case reviewed above. 

Then their expressions in the original radially moving frame \eqref{LineElement} can be obtained by performing the inverse coordinate transformation from the instantly rest frame with the special relativistic effect taken into account at each point. 
In order to fix the multiplicative factor, it is convenient to rewrite the local temperature in the instantaneously rest frame as $\check T = \check u^0$ up to a constant factor, where $\check u^0$ is the time component of the fluid velocity in the rest frame \cite{landau}.
This expression imposes the transformation rule of the temperature to be the same as that of time inducing its delay. This result matches the one given by Ott-Arzelis  \cite{Ott1963LorentzTransformationDW,Arzelis1965TransformationRD} and not by Planck-Einstein \cite{planckRT,einstein1907uber}.%
\footnote{Later, Einstein sent an informal letter to Laue suspecting that the correct transformation rule of the temperature is the one of Ott-Arzelis differently from his original paper \cite{Schröder_Treder_1994}. }
Therefore the local temperature $T$ in the current radially moving frame is given by $T=\check T/\sqrt{1-|\beta |^2}$, where $|\beta |=e^{\frac{B-A}2}|C|$ is the modulus of the velocity of the radially moving frame. 
This transformation rule is also the case to the thermodynamic local infinitesimal volume in order to be consistent with the local Euler's relation, $v=\check v/\sqrt{1-|\beta |^2}$.%
\footnote{This thermodynamic local infinitesimal volume transforms inversely to that of the infinitesimal interval of the space coordinate along the moving direction, which is consistent with the Fitzgerald-Lorentz contraction.}

How about other thermodynamic quantities? 
On the energy density and the pressure, recall that they can be defined directly through the energy-stress tensor. The energy density is $\rho= -T^0\!_0= \mathring\rho - 2 u\cd j + \varpi (u^r)^2$, the pressure in the radial direction $p= T^r\!_r =\mathring p+ \varpi (u^r)^2$ and that in any angular one $T^i\!_i = \mathring p$  in the original reference frame.
Note that this form of the energy density implies that $\varpi$ may be interpreted as the effective mass of the collective motion in the radial flow.  
The anisotropic pressure is consistent with the existence of flow momentum, and indeed the difference of the components of the pressure is related to the particle motion as $p-\mathring p=\varpi (u^r)^2$, which vanishes in the previous hydrostatic case as expected. 
Accordingly, there exists non-trivial particle number gradient in the radial direction and chemical potential $\mu$ for it. 
In order to determine the number density $\varkappa$ in the current radially moving frame, it is again convenient to recall it in the previous case of hydrostatic equilibrium, in which it was given by $\mathring\varkappa=\mathring j^0/\mathring u^0=\mathring j^0/\mathring T$, where a constant factor was absorbed suitably in the definition of $\varkappa$ and $\mathring j^0$ is the charge density in the rest frame. 
The index structure fixes the transformation rule of the charge density from the rest frame to the radially moving frame. As a result, the number density in the current radially moving frame is determined as $\varkappa= j^0/T$.%
\footnote{A caveat is that the current $j^\lambda$ is not given by the product of the number density and the flow velocity, $\varkappa u^\lambda$, in a general moving frame. This is because if $j^\lambda=\varkappa u^\lambda$, then it would give an extra undesirable constraint such that $j^0/u^0=j^r/u^r$ in the current radially moving frame.}
The chemical potential for the number density can be determined so as to satisfy the desired local thermodynamic relations. One is the local version of the Euler's relation $Ts = u + pv -\mu n$, where $s$ is the entropy density, $u=\rho v$ is the internal energy density and $n=\varkappa v$ is the thermodynamic number density with respect to the proper coordinates.\footnote{The concept of the density does not make sense in curved spacetime without specifying the volume element for the volume integration to obtain its total quantity. Here a density with respect to the proper coordinates means that its total quantity is obtained by the volume integration as usual in global thermodynamics.} The other is that of the Gibbs-Duhem relation $sdT = vdp - nd\mu$. Eliminating the entropy density $s$ from these two leads to $(u + pv -\mu n) d\log T = vdp - nd\mu$, or, $\varkappa T d(\mu/T) = dp -(\rho + p) d\log T$. 
On the other hand, using the thermodynamic variables, the relativistic fluid equation \eqref{ii} is rewritten as 
\beal{
p' = - \frac{ (d-2) }{r} (p -\mathring p ) +(\log T)'(\rho + p), 
\label{ii2}
}
or, $dp -(\rho + p) d\log T= - (d-2) (p -\mathring p ) dr/r$. Plugging this into the above thermodynamic relation yields $\varkappa T d(\mu/T) = - (d-2)\varpi (u^r)^2 dr/r$. Integrating both sides, one reaches the chemical potential as 
\be 
\mu =  - (d-2)T \int \frac{\varpi (u^r)^2}{j^0}\frac{ dr}r.
\label{ChemicalPotential}
\ee
Remark that $\mu/T$ is not constant any more in a gravitational equilibrium system in the presence of flow differently from the corresponding description in \cite{landau}. 
As a result the entropy density is determined through the local Euler's relation as $s =\frac{v}T( \rho + p + (d-2)  j^0 \int \frac{\varpi (u^r)^2}{j^0}\frac{ dr}r)$. Note that this formulation ensures the first law of thermodynamics as $Tds =du +pdv -\mu dn$.

Finally in order to derive the relativistic structure equations, consider a situation in which the spherically symmetric gravitational equilibrium system has boundary at a certain radius like a stellar object. For the purpose, it is standard to introduce the variable corresponding to gravitational mass or energy within the radius $r$ \cite{PhysRev.55.374}, which is denoted by $U_r$. It is not so trivial as in the hydrostatic case to specify $U_r$ since the metric becomes non-diagonal as \eqref{LineElement}. The proper definition turns out that $U_r =\frac{(d-2)}{2\kappa^2} \tilde V (1 - g^{rr})r^{d-3}$, where $\tilde V$ is the volume of the $(d-2)$-dimensional internal manifold given by $\tilde V=\int d^{d-2}x \sqrt{\tilde g}$.
Indeed, by employing the Einstein equation \eqref{0r} and \eqref{rr}, which have the following different expressions  
\beal{
&(d -2) \left( (g^{rr} -1) (d -3) - 2 r (\log T)' \right)  =2 \kappa ^2 r^2 p, 
\label{rr2}\\ 
&(g^{rr})'  =\kappa^2 \frac{2 r }{(d -2)} (-\rho - p ) - 2g^{rr} (\log T)'  , 
\label{0r2}
}
respectively, since $g^{rr}=e^{-B} - e^{-A}C^2, \varpi = (-\rho - p)/g^{rr}$, it can be shown by straightforward calculation that 
\be 
\frac{dU_r}{dr} =r^{d-2}\tilde V \rho. 
\label{UGradient}
\ee 
On the other hand, eliminating the metric component $g^{rr}$ from the definition of $U_r$ by using \eqref{rr2}, one finds $U_r =\frac{ \tilde V}{ 2\kappa^2 } \frac{-(d - 2) 2r(\log T)' - 2\kappa^2r^2p}{d - 3 - 2r(\log T)' }r^{d-3}$. Solving this with respect to the temperature gradient, one finds
\be 
\frac{dT}{dr} =-\frac{\kappa^2}{\tilde Vr^{d-2}} \frac{(d - 3) U_r + \tilde Vr^{d-1}p}{d - 2 - \frac{2\kappa^2 U_r}{\tilde Vr^{d-3}} }T.
\label{TGradient}
\ee
Plugging this into \eqref{ii2} leads to 
\beal{
\frac{dp}{dr} =- \frac{\kappa^2}{\tilde V}(\rho + p ) 
\frac{ (d - 3) U_r + \tilde Vr^{d-1}p} {r^{d-2}( d - 2 -\frac{2\kappa^2 U_r}{\tilde Vr^{d-3}}) }
- \frac{d - 2}r (p -\mathring p).
\label{pGradient}
}
This is the extended version of the TOV equation in the gravitational equilibrium system with steady radial flow, and the three equations \eqref{UGradient}, \eqref{TGradient}, \eqref{pGradient} are the general relativistic structure equations for the system.

Note that one variable $\mathring p$ increases due to the anisotropy of pressure generated by radial flow in the structure equations, and it is not constrained by the structure equations but by conditions for the current $j^\lambda$. 
A typical one is the covariant conservation equation but it may be more interesting to consider its anomalous situation generated by energy production. Such analysis of a model building by combining equations of state and possible boundary condition is beyond the scope of the paper and left to future work.   

\subsection{Poisson equation and heat conduction equation} 
\label{GRPE} 

For future applications, it is useful to derive the steady-state heat conduction equation from the structure equations. This can be done by simply eliminating variables suitably therefrom, though its concise form can be obtained by first deriving the general relativistic Poisson equation and converting it by employing the Tolman relation \cite{Yokoyama:2023gxf}. (See also \cite{Yokoyama:2024wad}.) 
For the purpose, it is required to specify the correct form of the gravitational potential.
What is the correct form of the gravitational potential in the current self-gravitating equilibrium system with the non-diagonal metric \eqref{LineElement}?

To answer the question, it is helpful to again consider a test particle with unit mass moving only in the radial direction of the gravitational system and make it probe the gravitational potential. If the test particle moves sufficiently slowly, then the probed gravitational potential $\phi$ must appear in the metric components as $g_{00}\approx -(1+2\phi)$ in the non-relativistic or weak gravity regime. On the other hand, in the static limit, there is an exact relation between the gravitational potential and the metric components as $g_{00}= -e^{2\phi}$ known as the Tolman relation \cite{PhysRev.35.904}. 
These imply that the local temperature is rewritten in terms of the gravitational potential $\phi$ as $T= T_\circ e^{-\phi}$, so that the Tolman relation of the equivalence between the local temperature and the gravitational potential holds precisely in the current steady system.

This argument and the specification of the gravitational potential can be verified explicitly by deriving the general relativistic Poisson equation on the assumption. 
The strategy is to compute $\nabla^2\phi$ in the background \eqref{LineElement} and rewrite it in terms of the energy-momentum tensor by using the equations of motion. Since $\phi$ is a scalar in the background, $\nabla^2\phi$ can be computed as $\nabla^2\phi=\frac1{\sqrt{-g}}\partial_\lambda \sqrt{-g}g^{\lambda\sigma} \partial_\sigma\phi =g^{rr} \{ \phi'' + \phi'^2 +( \frac{d-2}r + \half \frac{(g^{rr})'}{g^{rr}} ) \phi' \} $, where $(A+B)' =2\phi' - (g^{rr})'/g^{rr}$ was used. Differentiating both sides of \eqref{rr2} and substituting $(g^{rr})'$ by using \eqref{0r2}, one obtains 
\beal{ 
\phi '' =& \phi'^2 +  \frac{ 1 }{2 r g^{rr}}\left[ \phi' \{ - 2 r \left(g^{rr}\right)' -d +3 +\left(d - 5\right) g^{rr}-\frac{2 \kappa ^2 r^2 p}{(d -2)} \}  - (d -3) \left(g^{rr}\right)' + \frac{ 2 \kappa ^2 (r^2 p)'}{(d -2)} \right], \notag
}
where $\phi'=-(\log T)'$ was used. 
Substituting this into the above gives $\nabla^2\phi=2 g^{rr} \phi'^2 - \phi' \left(\half \left(g^{rr}\right)'+\frac{ \kappa ^2 r p}{d -2} + \frac{(d -3)}{2r}(-3 g^{rr}+1) \right)-\frac{(d -3) \left(g^{rr}\right)'}{2 r}+\frac{\kappa ^2 \left(r p'+2 p\right)}{d -2}$.  
Plugging \eqref{ii2} and \eqref{0r2} into this yields $\nabla^2\phi =\phi' ( g^{rr} \phi' - \frac{\kappa ^2 r p}{d -2}+\frac{(d -3) \left(g^{rr}-1\right)}{2 r})+\frac{\kappa ^2 ((d -2)\mathring p+(d -3) \rho+p)}{d -2}$, in which the first term vanishes by using \eqref{rr2}. As a result the general relativistic Poisson equation is obtained as 
$\nabla^2\phi =\frac{\kappa ^2}{d -2} ((d -3)\rho+p+(d -2)\mathring p)$, and this can be converted into the heat conduction equation by using the Tolman relation as $\nabla^2\log T =-\frac{\kappa ^2}{d -2} ((d -3)\rho+p+(d -2)\mathring p)$.
Note that these have been derived without using any approximation and thus holds exactly, and that they reduce to the previous results in the hydrostatic limit \cite{Yokoyama:2023gxf}. 

\section{Discussion}

It has been seen that turning on flow in a hydrostatic equilibrium system gives rise to various fruitful problems and physics. 
It is a priori unknown whether such a system reaches a steady state with flow,  and even though the answer is yes, it is a non-trivial mathematical problem to construct such an explicit gravitational solution even perturbatively, since it is required to specify the correct forms of the metric tensor and the energy-momentum one by adding terms suitably.
To address the problem, a physical intuition is useful as seen in this paper investigating the case to turn on radial flow without breaking spherical symmetry.
In this case it has been important to consider a matter current describing the flow and its contribution to the energy-momentum tensor, and the corresponding off-diagonal term in the metric tensor. 

The addition of an off-diagonal component of the metric induces the motion of the reference frame of coordinates, and in a general relativistic system, the velocity of the reference frame depends on each point of spacetime in general. 
In such a gravitational equilibrium system, it was not clarified how to define thermodynamic observables in accordance with the theory of relativity and the laws of thermodynamics. 
The paper has devised a new strategy to solve this problem. 
A focus is that thermodynamic observable quantity defined only through geometric quantities can be determined by reading them with an instantly rest frame taken and yielding the special relativistic effect pointwise induced by the inverse transformation to the original radially moving frame. 
The number density associated with the introduced current has been similarly determined by this prescription. 
On the other hand, the energy density and the pressure have been determined directly from the energy-stress tensor. By employing this result the thermodynamic internal energy density has been fixed as the product of the energy density and the infinitesimal volume. 
The validity of this identification of the thermodynamic observables has been verified by their admitting the desired local thermodynamic relations such as the local Euler's relation, the local Gibbs-Duhem one, and the first law of local thermodynamics. This ensures the existence of the entropy density and the chemical potential, and indeed their explicit forms have been derived. 
The procedure how to solve the problem presented here is expected to be applicable to another self-gravitating equilibrium steady system such as a rotating one of perfect fluid by modifying it suitably according to the situation. 

It would be interesting to construct the entropy current in this system with steady flow following the method in 
\cite{Aoki:2020nzm,Yokoyama:2023nld}. (See also \cite{Yokoyama:2023kkg}.)
It is clearly not given by $su^\lambda$ as is the case with $j^\lambda\not=\varkappa u^\lambda$, though such a description is often seen in textbooks and researches to analyze a system employing fluid dynamics.

A final goal of the program to investigate a self-gravitating equilibrium system including radial steady flow is to construct a stellar model with steady radial flow inside it to describe the radiation zone deep inside. Such radiation is supposed to be produced by the nuclear chain reaction near the core. It is important to set up equations of state which reflect the physical phenomena suitably. 

The progress of the program is hopefully reported in near future. 

\section*{Acknowledgement}
This work was supported in part by the Grant-in-Aid of the Japanese Ministry of Education, Sciences and Technology, Sports and Culture (MEXT) for Scientific Research (No.~JP22K03596). 

\appendix
\section{Ricci curvature tensor} 
\label{Ricci}

This appendix presents the result for Ricci curvature tensor for the line element \eqref{LineElement}.
\beal{
\Ric^{0}\!_{0} 
=& \frac{e^{-B -A }}{4 r} \bigg[2 e^{B } C  \left(C ' \left(2 d +3 r B '-r A '-4\right)+2 r C ''\right)-e^{A } A ' \left(2 d -r B '-4\right)\nn
& +e^{B } C ^2 \left(B ' \left(2 d -r A '-4\right)+2 r B ''+r B '^2\right)+4 r e^{B } C '^2-2 r e^{A } A ''-r e^{A } A '^2\bigg], \nn
\Ric^{0}\!_{r} 
=& \frac{(d -2) e^{-A } C  (B+A)'}{2 r} , \quad 
\Ric^{r}\!_{0} 
= 0, \nn
\Ric^{r}\!_{r} 
=& \frac{e^{-B -A }}{4 r} \bigg[-2 e^{B } C ^2 \left((d -2) A '-r B ''\right)+B ' \left(2 (d -2) e^{A }+r A ' \left(e^{A }-e^{B } C ^2\right)+6 r e^{B } C  C '\right) \nn
& +2 e^{B } C  \left(C ' \left(2 d -r A '-4\right)+2 r C ''\right)+r e^{B } C ^2 B '^2-r \left(-4 e^{B } C '^2+2 e^{A } A ''+e^{A } A '^2\right)\bigg], \nn
\Ric^{i}\!_{j} 
=& \frac{e^{-B -A }}{2 r^2} \bigg[e^{B } C ^2 \left(2 d +r B '-r A '-6\right)+e^{A } \left(2 (d -3) \left(e^{B }-1\right)+r B '-r A '\right)+4 r e^{B } C  C '\bigg] \delta^i_j. \notag
}
Throughout the paper, the prime means the differentiation with respect to the radial coordinate. 
From these it is straightforward to calculate other curvature tensor such as the Einstein tensor. 

\bibliography{RF}

\end{document}